\documentclass[aps,manuscript]{revtex4}
\usepackage[latin9]{luainputenc}
\usepackage{color}
\usepackage{float}
\usepackage{graphicx}

\makeatletter

\@ifundefined{textcolor}{}
{%
 \definecolor{BLACK}{gray}{0}
 \definecolor{WHITE}{gray}{1}
 \definecolor{RED}{rgb}{1,0,0}
 \definecolor{GREEN}{rgb}{0,1,0}
 \definecolor{BLUE}{rgb}{0,0,1}
 \definecolor{CYAN}{cmyk}{1,0,0,0}
 \definecolor{MAGENTA}{cmyk}{0,1,0,0}
 \definecolor{YELLOW}{cmyk}{0,0,1,0}
}

\makeatother

\begin{document}
\title{Three-level multilevel simulation of the neuroreceptor function of
NMDA}
\author{Francesco Di Palma$^{1}$, Sauro Succi$^{1}$, Fabio Sterpone$^{2}$,
Marco Lauricella$^{3}$, Franck Perot$^{4,6}$, Simone Melchionna$^{4,5}$}
\affiliation{$^{1}$Istituto Italiano di Tecnologia, Central Research Labs, Via
Morego 30, 16163 Genova, Italy}
\affiliation{$^{2}$Laboratoire de Biochimie Th\'eorique, UPR 9080, CNRS Universit\'e
Denis Diderot, Sorbonne Paris Cit\'e, IBPC, Paris, France}
\affiliation{$^{3}$Istituto Applicazioni Calcolo M. Picone CNR, Via dei Taurini
19, 00185 Rome, Italy}
\affiliation{$^{4}$Istituto Sistemi Complessi CNR, Via dei Taurini 19, 00185 Rome,
Italy}
\affiliation{$^{5}$MedLea Srls, Italy}
\affiliation{$^{6}$Lexma Technology LLC, USA}
\begin{abstract}
We present a new multiscale method to study the N-Methyl-D-Aspartate
(NMDA) neuroreceptor starting from the reconstruction of its crystallographic
structure. Thanks to the combination of homology modeling, Molecular
Dynamics and Lattice Boltzmann simulations, we analyse the allosteric
transition of NDMA upon ligand binding and compute the receptor response
to ionic passage across the membrane. 
\end{abstract}
\maketitle

\section{Introduction}

Modern biological sciences are incessantly searching to unveil the
molecular mechanisms of biological activity, their dis-functionalities
and the ensuing diseases so as propose and develop possible remedies.
Computer models have a major say in this endeavour, as they can often
access regions of parameter space which are beyond reach of direct
experimentation. In particular, molecular dynamics simulations have
a long and time-honored history in the study of biomolecular assemblies
culminating with the 2013 Nobel Prize in Chemistry \citet{nobelchemistry}.

The field shows no sign of flagging; quite on the contrary it is witnessing
burgeoning progress in multiple directions, and particular in terms
of capabilities to tackle larger systems over longer temporal scales.
Remarkable examples in point are the recent simulations of biological
systems is the 100-million atom-scale model of an entire cell organelle,
a photosynthetic chromatophore vesicle from a purple bacterium \citet{rochaix2019dynamic}
or the study of the N-Methyl-D-Aspartate (NMDA) neuroreceptor by the
DE Shaw research group \citet{song2018mechanism}. These outstanding
achievements were obtained by exploiting the method of Molecular Dynamics
(MD), whereby all atoms in the simulation are taken into account,
with a massive computational effort, often undertaken on special hardware. 

Such achievements unquestionably push the frontiers of simulation
further ahead and pave the way to disruptive progress in many areas
of biology and medicine. In particular, the possibility of simulating
large systems is key to neuroscience, where the response of neuro-receptors
to binding with natural or synthetic ligands can reveal the allosteric
transitions in the membrane protein and how the channel pore function,
with the modulation of the ionic passage. NMDA receptors, in particular,
play a critical role in brain development and function, including
learning and memory formation \citet{trimble2002molecular,burnell2018positive}.
Dysfunctional NMDA is implicated in various neurological disorders,
such as Alzheimer's disease, depression, stroke, epilepsy and schizophrenia.
Therefore, computational capability to tackle large-scale protein
transitions and ion passage has the potential to study many neurological
diseases and design new drugs through medicinal chemistry methods.
Some initial studies have indeed tackled the study of NMDA by a brute
force MD approach that, although showing some degree of success, has
made clear the difficulty in covering the complete times scale of
the allosteric and functional response \citet{dai2013nmda,zheng2017probing,song2018mechanism,palmai2018does}. 

Under physiological conditions, the opening of the receptor ion channel
requires concurrent binding of glycine and glutamate and relief of
magnesium block at the ion channel pore by membrane depolarization.
The resulting calcium flux triggers a cascade of signal transduction
necessary for synaptic plasticity. 

Multiscale methods offer a strategic advantage in simulating such
receptors that lay at the basis of many basic neurological phenomena.
The reasons is that neuroreceptors are large molecular assemblies
with large stretches of membranes separating intra and extracellular
regions and receptors (proteins) as gates. Binding small molecules
(neurotransmitters) triggers spatially small but functionally very
important structural changes (allosteric transitions) that open/close
the gates to the passage of ions. 

From the neurological viewpoint, it is key to quantify the allosteric
response to the binding of nanomolar quantities of neurotransmitters.
These can be natural or artificial, with important applications to
medicinal chemistry. Once the allostery is measured and quantified,
one can measure passage of ions in the receptor pore. To this purpose,
electrokinetics raises a significant challenge, as it implies the
presence of blocking elements (Magnesium for NMDA), since the process
is an activated one. To achieve such ambitious program all-atoms simulations
fall short of covering the wide stretch of spatial and temporal scales
involved, ranging from picoseconds to minutes and with a number of
atomic degrees of freedom in excess of millions. In essence, reproducing
binding affinity (thermodynamics), allosteric transitions (large-scale,
fine and coarse atomic motion) and electro-kinetics, is a task that
involves too widely disparate scales to be liable to a single level
of representation and simulation. 

For these reasons, we have devised a new (three-level) multiscale
method to study neuroreceptors starting from the reconstruction of
the protein structure, as obtained from incomplete crystallographic
data, to analyze the binding affinity of NMDA to its ligands, followed
by the quantification of the allosteric motion, and finally to assess
the modulation of the ionic passage. 

Our approach begins by using homology modeling techniques to reconstruct
the NMDA configuration in the phospholipidic environment, the simulation
of the allosteric events by means of MD and finally by the study of
the electrokinetic ion transport by means of a Lattice Boltzmann approach.\textcolor{black}{{}
The novelty of our approach relies on the use of LB electrokinetics
to analyse the flow of ions in a probabilistic way. While other strategy
can be powerful to study the passage of single or multiple charges
in ion channels via the potential of mean force method \citet{roux2004theoretical}
or in a statistical sense, by solving the Poisson-Nernst-Planck equation
\citet{kurnikova1999lattice}, our approach takes into account the
complete hydrodynamic modes of the saline solution flowing through
the receptor matrix. In this way it is possible to observe direcly
the effect of electrostatic forces and convection on the three-component
solution made of neutral, cationic and anionic species. The approach
thus allows selecting a large set of protein configurations so that
the prediction of flux based on configuration allows direct comparative
analysis with purely steric approaches \citet{smart1996hole}. Finally,
the proposed method can be enriched by combining the sampling of configurational
space with adequate simulations of all atoms or coarse grained models,
in order to follow the flow in different conditions. }

The multiscale and multilevel approach is entirely novel in concept
and application and demonstrates, for the first time to the best of
our knowledge, that complex biological processes occurring in neuroscience
can be tackled, thus providing a whole wealth of microscopic, dynamic
and functional information that is unaccessible to current experimental
techniques. 

\section{Computational approach}

\begin{figure}[H]
\begin{centering}
\includegraphics[scale=0.7]{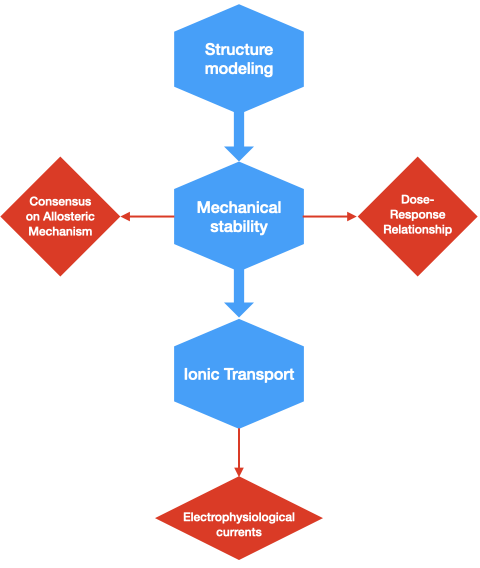}
\par\end{centering}
\caption{Protocol Flowchart illustrating the three-level approach to studying
the neuroreceptor function. Top: starting from the incomplete crystallographic
data, homology and structure modeling allows reconstructing the initial
condition. Middle: by applying MD all-atom and coarse grained simulations
one can access the allosteric transition and compute affinity to binding,
allowing to determine the dose-response relation. Finally, use of
the fluctuating protein matrix enables the use of LB to quantify the
eletrokinetic content of the receptor, thereby constructing the dose-function
relation.}
\label{fig:Flowchart}
\end{figure}

As shown in Fig \ref{fig:Flowchart}, our computational approach involves
a highly orchestrated network of computational methods and software
tools, which can broadly classified within three basic levels:
\begin{enumerate}
\item Structure modelling (Homology Methods); 
\item Mechanical stability (Molecular Dynamics and Molecular Mechanics);
\item Ionic transport (Lattice Boltzmann).
\end{enumerate}
The whole three-level procedure spans about six decades in time, from
nanoseconds up to milliseconds, reaching up to timescales of physiological
relevance. The ultimate outcome of clinical relevance is the dose
response relationship, namely the current flowing across the neurotransmitter
as a function of the dose of ligand.

To accomplish the three segments of the flowchart, we used three different
softwares: Modeler \citet{eswar2006comparative} for Homology Modeling
and Gromacs \citet{abraham2015gromacs} for All-Atoms MD. Most of
simulations are based on the software Moebius from Lexma Technology,
as it allows multiphysics simulations of multicomponent fluids and
particles. Therefore, Moebius was then used for the Coarse-Grained
MD and for the Lattice Boltzmann simulations. We utilized both the
CPU and GPU architectures to accomplish the heavy-computing runs,
as available on both the Gromacs and Moebius softwares.

\subsection{Structure Modeling}

The first level of the flowchart pertains to characterising the molecular
structure of the neuroreceptor, consisting of the protein embedded
in the cellular membrane. Homology modelling allows to construct the
atomic-resolution model of NMDA GluN1A/GluN2B starting from its amino
acid sequence and the available crystallographic three-dimensional
structure (PDB id: 4PE5) \citet{karakas2014crystal}. For crystallographic
data from membrane proteins with a large amount of missing structures,
such as for NMDA, is rather challenging for homology modeling. 

Reconstructing the missing secondary structure is accomplished by
resorting to homological techniques whereby the atomic-resolution
model of the target protein from its amino acid sequence is constructed
by comparing the experimental three-dimensional structure of a template
homologous protein. The reconstruction relies on the identification
of known secondary structures from homologous receptors acting as
templates that resemble the structure of the query sequence and on
an alignment that maps residues to residues in the template sequence.
The identification of one or more known structures resembling the
one of the query sequence and the alignment of residues in the query
sequence to residues in the template sequence, benefits from the fact
that NMDA structure is conserved amongst homologues.

\subsection{Mechanical Stability and Allostery}

The second level of the flowchart pertains to assessing the stability
of the aforementioned molecular complex, including its interactions
with the water solvent, and assessing the allosteric transition. Once
ligands bind to the specific receptor sites, NMDA changes its shape,
and altering the affinity for a ligand at a second site (e.g., either
a receptor or a binding site); the ability of an effector molecule
(ligand) to change the conformation and activity of a protein.

In allostery, the function of a receptor is modified by the interaction
with its ligands, not only at the active site but also at a spatially
distinct site of different specificity. In allostery, the interaction
of the functional sites results in an altered affinity of ligand binding,
thus depending on the dynamic interaction with the substrate.\textcolor{black}{{}
Ideally the conformational changes induced by the binding of the allosteric
effector can be finely followed by brute force MD simulations at atomistic
resolution. For instance, if the allosteric response path is of interest,
starting from the substrate free equilibrated structure it is possible
to introduce the effector and follow in time its conformational changes
and fluctuations. This is the strategy used in our approach. In other
situations, where both the APO and HOLO states are available, the
thermal fluctuations in each state, and possible state interconversion
can be simulated with brute force MD or by enhanced sampling techniques
(e.g. parallel tempering, thermodynamic integration, metadynamics,
etc.). However, the high resolution of the full atomistic approach
is often limiting for very large systems. It is convenient therefore
to use a simplified coarse-grained model. This is particularly challenging
for membrane proteins that, for a proper treatment of allosteric process,
an adequate level of controlled flexibility of the protein scaffold
is required. In our approach we tested the capability of the OPEP
CG force field \citet{chebaro2012,sterpone2014,sterpone.opep.2013}
to reproduce the main fluctuations and conformational changes associated
to the allosteric transition by comparing the results with atomistic
modelling.}

\textcolor{black}{The AA protocols is as follows: we simulated both
the APO (ligand free) and the HOLO (ligand bound) versions of NMDA.
For the HOLO version, we positioned two Glutamate (GLU) and two Glycine
(GLY) neurotransmitters molecules in the NMDA pockets of the relaxed
model. This was obtained by docking them by means of the Chimera software
chain-by-chain superimposition to achieve suboptimal overlap of the
protein domains.}

\textcolor{black}{The OPEP force field has been designed to model
peptide and protein folding and aggregation in solution. It is a multi-scale
model that reserves an atomistic description to the backbone atoms
and reduces the side chain amino acid to a unique bead. OPEP was designed,
and progressively improved, for modelling soluble proteins \citet{chebaro2012,sterpone2014,sterpone.opep.2013}
but its coupling with an explicit membrane environment is under way. }

\textcolor{black}{Within the scope of the present investigation, we
adopted the available version with an adequate strategy to model the
trans-membrane part of the protein. The test was performed on two
models. In the first one the protein motion is fully flexible and
its motion is controlled by the OPEP hamiltonian. Some restrains are
applied to the trans-membrane part of the proteins so to mimic the
spatial embedding of the membrane. The flexibility is key for monitoring
the loops motions in the extra-membrane regions that control the allosteric
response. The second model uses a more radical simplification, and
is based on a floppy elastic network of the protein. On top of the
elastic network, non-bonded sites interact via the OPEP non-bonded
potential. To be noted that in several previous works the characterisation
of protein allosteric paths are modelled by considering elastic networks
\citet{rocks2017designing}. In the CG approach for the allosteric
effectors (GLN and GLY amino acids) instead we used the same OPEP
model but their location in the binding site was restrained. It is
worth noting that the OPEP force field was already successfully coupled
to the hydrodynamic description based on LB \citet{sterpone2015},
allowing the investigation of complex processes like amyloid aggregation
\citet{chiricotto.jcp.2016,chiricotto.jpcl.2019}, protein crowding
\citet{sterpone2014,timr2020} and unfolding under shear flow \citet{sterpone.shear.2018,cattoen2018}.}

\subsection{Ionic Currents}

Finally, the third level pertains to simulating the ionic currents
within the neuroreceptor, which is performed in order to quantify
the conductivity of the neuroreceptor as a function of the applied
voltage or salinity and for varying configurations of the receptor.
The Lattice Boltzmann (LB) method provides a particularly useful approach
to investigate the ionic response since it is rooted in the mesoscopic
description of matter which perfectly maps the current representation
of the receptor and its environment \citet{succibook}. 

LB is a very convenient computational approach since it is based on
a over a Cartesian grid, that is, by using a uniform mesh based on
cubic voxels. It is highly adaptable to reproduce the flow structure
of single and multi-component system under the action of external
or internal forces. In particular, LB has been shown in the past to
be capable of reproducing the electro-kinetics of saline solutions
in model devices with non-trivial geometries and locally charged surfaces
\citet{MARCONI2011,melchionna2011electro}. 

The case of the NMDA receptor, however, presents a non trivial geometry
within the pore region and highly localized forces $F(x)$ at position
$x$ stemming from charged or neutral atomic groups. In addition,
the local electrostatic forces can be locally intense, often exhibiting
rapid spatial modulations of the electrolytic densities due to the
formation of disordered double layers. Such scenario poses several
challenges to the computational scheme, particularly by endangering
numerical stability as dictated by the Courant-Friedrich-Lewy stability
condition $F(x)\lesssim m\Delta x/\Delta t^{2}$. It is well-known
that the presence of unit charges in simulation cannot be simulated
by a direct LB approach \citet{raafatnia2014computing} without proper
treatment to enhance its stability.

To circument such limitations, we employed a multi-component Entropic
Lattice Boltzmann Method, a powerful variant of the basic Lattice
Boltzmann method based on a self-consistent tuning of the relaxation
parameter so as to ensure compliance with local entropy growth (H
theorem). 

The LB approach to simulating the saline solution is based on tracking
the evolution of each fluid component, with the index $\alpha=0,1,2$
labelling the neutral aqueous medium ($0$), and the positively ($1$)
and negatively charged components ($2$), via the discretized form
of the density distribution function, named populations $f_{p}^{\alpha}(x,t)$.
Here, $t$ denotes the temporal coordinate and subscript $p$ labels
a set of discrete speeds $c_{p}$ connecting the mesh points to its
neighbors. The two ionic components are monovalent and characterised
by charges density $n^{\alpha}$ and velocity $u^{\alpha}$ and, given
the barycentric velocity $u=\frac{\sum_{\alpha}n^{\alpha}u^{\alpha}}{\sum_{\alpha}n^{\alpha}}$,
the relative velocity being denoted $\delta u^{\alpha}=(u^{\alpha}-u)$.
The neuroreceptor and the membrane are described by the collection
of particles at position $\vec{r}_{i}$ being neutral or partially
charged with charge $q_{n}$ valence $z_{n}$ provided by the force
field.

In standard multicomponent LB the dynamics of each component then
follows its own evolution equation:

\begin{equation}
f_{p}^{\alpha}(r+c_{p},t+1)=f_{p}^{\alpha}(r,t)+\omega(f_{p}^{\alpha,eq}-f_{p}^{\alpha})+S_{p}^{\alpha}\label{eq:LBMmulti}
\end{equation}
where $\omega$ is a relaxation frequency related to the kinematic
viscosity as $\nu=c^{2}\left(\frac{1}{\omega}-\frac{1}{2}\right)$,
$c=1/\sqrt{3}$ the lattice speed of sound, and the Maxwellian equilibrium
given by \citet{MARCONI2011,MARCONI2011b},

\begin{equation}
f_{p}^{\alpha,eq}=w_{p}n^{\alpha}\left[1+\frac{\delta u^{\alpha}\cdot c_{p}}{c^{2}}+\frac{(\delta u^{\alpha}\cdot c_{p})^{2}-c^{2}(\delta u^{\alpha})^{2})}{2c^{4}}\right]\label{eq:equilibrium}
\end{equation}
and $w_{p}$ is a set of normalized weights. The force term is $S_{p}^{\alpha}=w_{p}n^{\alpha}\left[\frac{F^{\alpha}\cdot c_{p}}{c^{2}}+\frac{(c_{p}\cdot u)(c_{p}\cdot F^{\alpha})-c^{2}F^{\alpha}\cdot u)}{c^{4}}\right]$
and the local force $F^{\alpha}=-ez^{\alpha}\mathbf{\nabla}\psi-\omega_{drag}^{\alpha}\sum_{\beta}\frac{n^{\beta}(u^{\alpha}-u^{\beta})}{n}+\gamma u^{\alpha}\tilde{\delta}(r_{n}-x)$,
being the sum of the self-consistent electrostatic forces, the inter-specie
drag force characterized by the cross-diffusion coefficient $D^{\alpha}=c^{2}/\omega_{drag}^{\alpha}$,
and the frictional force exerted by the receptor and membrane atoms
on the fluid species, being proportional to the coefficient $\gamma$.
In addition, $\tilde{\delta}(x)$ is a function used to smear the
particle charge on the LB grid, as used in Immersed Boundary method
\citet{peskin2002immersed}.

The electrostatic potential $\psi$ is the solution of the Poisson
equation, $\nabla^{2}\psi=-\frac{1}{\epsilon}[en^{+}(x)-en^{-}(x)+\sum_{n}z_{n}q_{n}\tilde{\delta}(r_{n}-x)]$
in the aqueous medium of dielectric permittivity $\epsilon$, $e$
being the unit electronic charge. In this study the LB solution is
obtained on a cartesian grid of spacing $0.01$ nm, by employing the
D3Q19 set of discrete speeds and associated weights, and by smearing
particle charges over $2\times2\times2$ grid points. We chose a ionic
cross-diffusivity of $2$ nm$^{-2}$/ns and a frictional coefficient
$\gamma=0.5$ ns$^{-1}$.

The complete multicomponent Entropic LB evolution is obtained by solving
eq. \ref{eq:LBMmulti} complemented by the minimization of the lattice
H-function $H^{\alpha}[f^{\alpha}]=\sum_{p}f_{p}^{\alpha}log(f_{p}^{\alpha}/w_{p})$,
which additionally provides a small local adjustment to the relaxation
frequency $\omega$ to enforce stability. Under operating conditions
we found that applying a filtering approach to populations to remove
non-hydrodynamic modes \citet{kramer2019pseudoentropic} significantly
enhances numerical stability \citet{montessori2014regularized}, providing
the final conditions to simulate saline solutions in presence of unit
charges. 

\section{Results}

In the following, we provide a summary of the main results of our
analysis. 

\subsection{Homology Modeling}

We first considered the tetrameric NMDA GluN1A/GluN2B structure from
\emph{Rattus Norvegicus} reconstructed by Homology Modeling at atomic-resolution
model starting from the three-dimensional crystallographic structure
\citet{karakas2014crystal}.

The structure lacks several loops in the linker region between the
ligand-binding domain and the missing loops were reconstructed chain-by-chain
using the high precision DOPE-HR modeling protocol \citet{shen2006statistical},
resulting in the most accurate available refinement method to obtain
high quality structural stretches. 

Subsequently, the receptor was relaxed using an energy minimization
scheme first \emph{in vacuum} by means of MD via a steepest descent
algorithm to overcome/solve bad contacts and improve the overall quality
of the structure quality. 

Validation of both post-homology model and equilibrated structures
was obtained by using MolProbity \citet{williams2018molprobity}.
Once relaxed, the receptor was embedded in a phospholipidic membrane
and prepared for simulation by surrounding the entire system in water
molecules at ambient temperature and as density corresponding to $1000$
Kg/m$^{3}$. 

\subsection{Molecular Dynamics and Molecular Mechanics simulations}

\textcolor{black}{Once the system was validated, we performed MD simulations
at three increasing levels of coarse graining, namely }\textcolor{black}{\emph{i)}}\textcolor{black}{{}
All-atom resolution based on CHARMM36m force field, }\textcolor{black}{\emph{ii)}}\textcolor{black}{{}
Coarse-grained MD based on OPEP force field, }\textcolor{black}{\emph{iii)}}\textcolor{black}{{}
elastic network representation. }We simulated both the APO (ligand
free) and the HOLO (ligand bound) versions of NMDA. For the HOLO version,
we positioned two Glutamate (GLU) and two Glycine (GLY) neurotransmitters
molecules in the NMDA pockets of the relaxed model. This was obtained
by docking them by means of the Chimera software chain-by-chain superimposition
to achieve anoptimal overlap of the protein domains. 

At first the all-atom resolution model was chosen by employing the
CHARMM36m force field \citet{huang2013charmm36} to represent membrane
proteins by taking into account bonding and non-bonding interactions.
The former includes bonding, angular and torsional forces among peptidic
groups while the latter include Van der Waals, electrostatic and hydrogen
bond forces. 

\textcolor{black}{The atomistic simulations successfully model the
allosteric effect, since upon the binding of the effector (GLN or
GLY) we observe the opening of the channel associated to the specific
reorganization of the extra-membrane region. This opening is observed
in the timescale of $300$ ns.}

A first-level coarse-grained model was utilized by employing the OPEP
force field \citet{sterpone2018multi} to represent amino acids fully
explicitly for the peptidic backbone, inclusive of hydrogen atoms,
while the lateral groups are represented by an effective particle.
Such representation is particularly accurate for the backbone as it
includes stretching, angular and torsional movements together with
non-covalent and hydrogen bonds. The pseudo-particles for the lateral
groups take into account the sterical and non-covalent interactions,
together with explicit representations of the saline bridges. 

\textcolor{black}{The CG simulations were performed and contrasted
against the atomistic one taken as reference. We observed a comparable
opening of the channel. The result supports the use of this less time
consuming model to generate a valid ensemble of configurations for
further electrokinetic analysis. A detailed description of the trajectories
and the conformational motions associated to the allosteric transition
is reserved to a further work.}

A further level of coarse graining was employed by using an elastic
network representation. Here all intramolecular bonding and non-covalent
forces were substituted with harmonic interactions that allow for
a certain level of protein deformability. Intermolecular forces are
still accounted for by means of the OPEP force field. 

All three models showed to undergo the allosteric transition once
the ligands have been positioned in the corresponding pockets. 

The transition took place on the $100$ ns timescale and several attempts
were observed before NMDA finally reached its stable configuration,
which is supposed to be an open pore configuration. Importantely,
the consensus of the three models to reproduce the transition lends
a significant degree of confidence to the overall homology-based reconstruction
and simulation models.

With the three representations of the neuroreceptor in place, we were
able not only to generate a large number of configurations with the
all-atom method, but also to harvest the relaxed structures by replacing
the membrane and receptor by order two different levels of coarse
graining, the OPEP and elastic network models. 

\subsection{Ionic passage}

A purely steric analysis of the pore geometry is initially performed
by finding the best route for a sphere with variable radius to squeeze
through the channel. The method HOLE allows the analysis of the dimensions
of the pore running through a structural model of an ion channel \citet{smart1996hole}
where the algorithm uses a Monte Carlo simulated annealing procedure.
The method predicts conductance by using a simple empirically corrected
ohmic model. However, ion permeation cannot be simply identified from
the physical dimensions of the pore. For example, water within narrow
hydrophobic pores can modulate permeation without even requiring steric
occlusion of the pathway. Better methods have been proposed to account
for hydrophobic gating, such as in the CHAP method \citet{klesse2019chap}. 

\begin{figure}[H]
\begin{centering}
\includegraphics[scale=0.50]{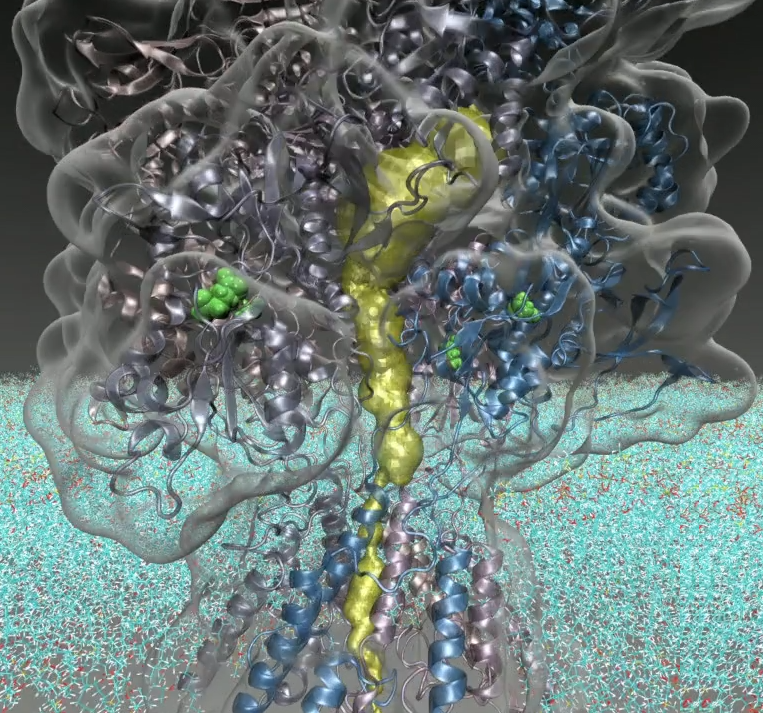}
\par\end{centering}
\medskip{}

\caption{Example of passage obtained with the HOLE method exemplifying one
possible conduction pathway and its narrow steric space.}
\label{fig:holeExample}
\end{figure}

\begin{figure}[H]
\begin{centering}
\includegraphics[scale=0.25]{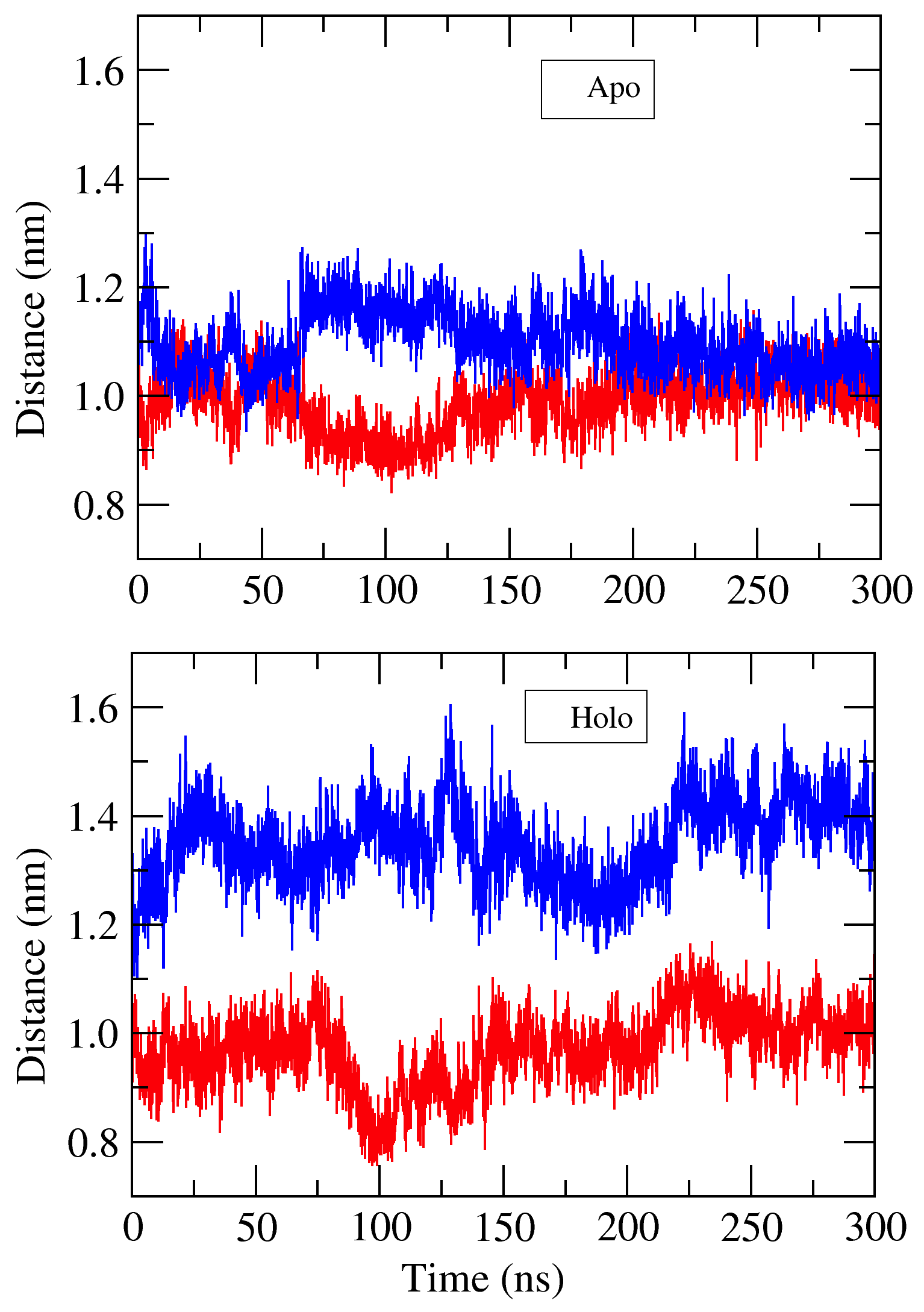}
\par\end{centering}
\medskip{}

\caption{Time evolution of the two gating bottleneck distances in the NMDA
pore for the APO (upper) and HOLO (lower) states. In both panels,
in red the distance between the CoM of V632 and V2244 of the two GluN1A
monomers, in blue the distance between I1426 and I3039 of the GluN2B
monomers.}
\label{fig:Gating Bottleneck}
\end{figure}

The effective passage of ions can take place in any of the small apertures
and crevices present within the receptor matrix, that can be seen
as an effective porous medium. Fig. \ref{fig:holeExample} illustrates
the steric passage within NMDA for a given receptor conformation.
The available volume is rather funnel shaped and the receptor bottleneck
is very narrow, with a lateral aperture being smaller than one Angstrom.
By analysing the time evolution of the bottlenck distance, computed
as the distance between the two GluN1A subunits considering the CoM
of Val632 and Val2244, and between the GluN2B subunits Ile1426 and
Ile3039, as reported in Fig. \ref{fig:Gating Bottleneck}, we observe
a differentiated behavior between the APO and HOLO states, suggesting
that the allosteric transition has taken place. In accordance with
the all-atom simulations, a similar behaviour has been observed analysing
the bottleneck distances in the CG and Elastic Network simulations
(data not shown); thus resulting in a coherent representation of the
system allosteric movement under the influence of the ligands, for
the three difference models. However, such narrow space renders a
quantification of the total ionic resistance highly dependent on the
multiple bottlenecks and conduction channels along the pathway. 

The presence of small fluctuations and the presence of several charged
groups along the pathway, renders the purely steric analysis qualitative
and even hard to justify. Upon binding of the receptor with the ligands,
the allosteric modification alters ionic resistance in several ways:
the variations in the electrostatic environment along the pore extension,
the fluctuating motion of the protein matrix in the channel, the presence
of water that is advected and that alters the hydrophobic content
of the pore, the sub-Angstrom modification of the pore bottlenecks.
These conditions can result in modifications of the barrier crossing
rates by orders of magnitude. Once again, the complex interplay of
these conditions and their associated timescales rules out a computational
approach based on a direct atomistic approach since it would require
hundreds of nanoseconds of observation time for monitoring a single
ion passage. Consequently, gathering sufficient statistics is out
of reach. The proposed scheme instead provides an effective decoupling
between the macromolecular motion and the ionic passage, thereby allowing
to study the motion of the saline solution during the macromolecular
evolution.

The simulation on ionic transport proceeds as follows. MD simulations
deliver a time sequence of structural configurations of the neuroreceptor,
$\lbrace\mathcal{C}_{t},\;t=0,t_{sim}\rbrace$. For a prescribed sequence
of such configurations, a long-time (hundreds of nanoseconds) LB simulation
is performed until the steady-state current, $J_{t}\equiv J_{ss}(\mathcal{C}_{t})$,
supported by the given configuration at time $t$ is obtained. In
passing, we note that this procedure also permits to accumulate significant
statistics, due to the the fact that the structural changes of NMDA
are pretty slow on the scale of the MD integration. The resulting
current $J_{t}$ shows abrupt up and down changes in time, which we
associate with opening (closing) of conductive channels within the
receptor configuration. Needless to say, the structural dynamics of
these channels is extremely rich, with abrupt morphological changes,
such as sudden strictions which eventually quench an otherwise highly
conductive channel and viceversa. 

Functional response is quantified by considering both the APO and
HOLO versions of NMDA and considering a single $300$ ns MD trajectory.
Conformational analysis of the receptors and application of the HOLE
method exhibit large fluctuations of the pore region by steric analysis,
indicating that allosteric transition is effectively taking place
(data not shown). Conductance was measured by extracting $100$ protein
conformations, evenly spaced in time of the last $100$ ns of the
all atom simulations of the equilibrated receptor for the putative
close and open states. 

\begin{figure}[H]
\begin{centering}
\includegraphics[scale=0.3]{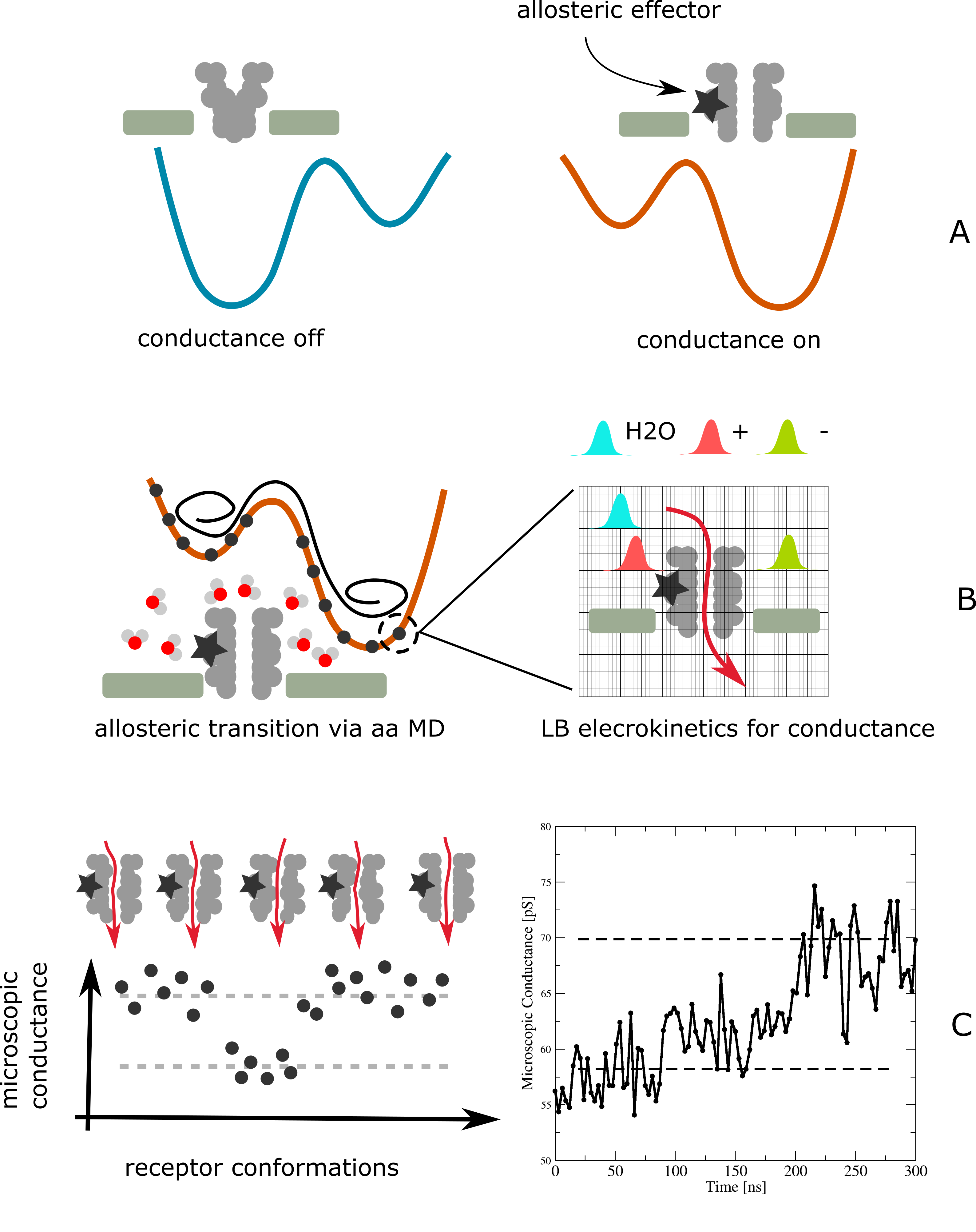}
\par\end{centering}
\caption{Representation of NMDA evolution upon binding with the ligands and
the computational scheme. A: the receptor evolves from the closed
to the open state, characterizaed by a change in conductance from
a baseline value to the one associated with the conducting state.
B: MD simulations sample the allosteric evolution of the receptor
conformations, characterized by large fluctuations during the $300$
ns single MD trajectory. C: by selecting configurations every $3$
ns after binding to ligands, the evolution of conductance is obtained
from LB simulations of a saline solution (made of neutral, anionic
and cationi species), whose data are shown in the rightmost column.
The horizontal lines are guides to the eye.}
\label{fig:three-levels-scheme}
\end{figure}

Figure \ref{fig:three-levels-scheme}C shows the conductance (in picoSiemens)
as a function of time. From this figure, several bursts of conductance
are visible past the open-up event mark the genuine microscopic nature
of conductance, driven by the underlyking fluctuations of the protein
matrix. These fluctuations provide a neat signature of channel opening/closing
events but with a clear trend towards stabilization of the open configuration.
It is expected that at longer times, not covered by the present simulations,
the open state will likely exhibit a consistent amount of short-lived
closed configurations.

Summarizing, within our procedure, the neuroreceptor is treated as
a slow time-changing molecular porous medium, and the geometrical
flexibility of LB is leveraged to compute the ``electrical permeability''
of such molecular porous media ``on-the-fly'' \citet{succi1989three}.

\subsection{Summary}

We have presented a new three-level multiscale method for the computational
study of the NMDA neuroreceptor.

The computational framework combines three distinct representation
levels:

i) Homology models for the characterisation of the protein structure,
ii) various forms of Molecular Dynamics for the dynamical stabili\textcolor{black}{ty
of the NMDA complex and finally iii) lattice Boltzmann simulations
of ion transport across the neurotransmitter. The novelty of our approach
relies on LB electrokinetics to analyse the mesoscopic flow of ions
by taking into account the complete hydrodynamic modes of the saline
solution as it flows inside the receptor matrix. It is thus possible
to observe direcly the effect of electrostatic forces and convection
on ionic currents at varying receptor configurational states. To this
aim, the usage of mesoscopic representation of the protein via a coarse
grained force field (such as OPEP used in this work) with the saline
solution described by the Boltzmann picture provides equivalent levels
of detail. In more general terms, the three-level multiscale method
described here can be further extended to combine the evolution of
the lagrangian representation of one subsystem, typically simulated
via MD, with the eulerian representation of a complementary subsystem,
simulated via LB, thereby taking into account the dynamical two-way
exchange of forces. }

The numerical results on NMDA show clear evidence of allosteric transitions
stimulated by binding of Glutamate and Glycine ligands.\textcolor{red}{{}
}Ionic transport across the NMDA complex show preliminary agreement
with experimental data. Importantly, the provided scheme makes the
study of the neuroreceptor functional response viable thanks to the
high performances of the MD and LB components. In particular the MD
component requires approx. 4 GPU hours per nanosecond, while the LB
requires 1 CPU/hour per simulation of electrokinetics, that translates
to 0.1 GPU/hour when running on a single GPU.

It is hoped and expected that the present three-level framework may
pave the way to the computational study of a variety of fundamental
multiscale biological processes. As presented in the text, such outcome
compares very favourably with single-channel patch clamps, thus enabling
future use of the present computational methodology to other types
of neuroreceptors. 

\section*{Acknowledgements}

This paper is dedicated to Mike Klein, a pioneering and inspiring
figure of molecular simulations for many decades.

SS wishes to acknowledge funding from the European Research Council
under the Horizon 2020 Programme Grant Agreement n. 739964 (``COPMAT''). 

\pagebreak{}

\pagebreak{}
\begin{figure}[H]
\begin{centering}
\includegraphics[scale=0.5]{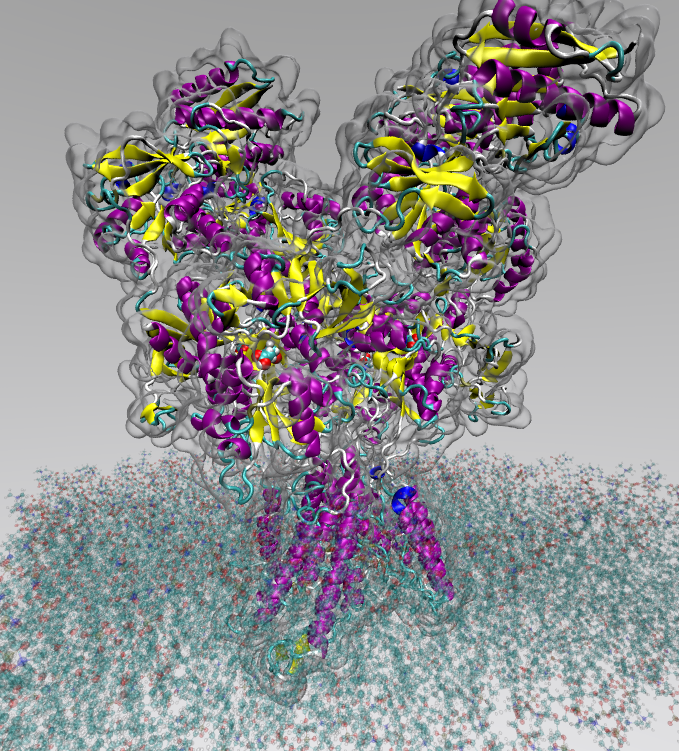}
\par\end{centering}
\medskip{}

\caption{Reconstructed NMDA receptor embedded in the POPC membrane and with
the GLN and GLY ligands. The water solvent has been removed from the
visualization for the sake of clarity.}
\label{fig:reconstructed}
\end{figure}

\pagebreak{}

\pagebreak{}


\paragraph{\bibliographystyle{plain}
\bibliography{lbmd}
}
\end{document}